\documentclass[journal, 11pt]{IEEEtran}
\usepackage{geometry}

\usepackage{amssymb,color}
\usepackage{graphicx}
\usepackage{physics}
\usepackage{url}
\urldef{\mailsa}\path|{tz636, qz494}@nyu.edu|

\let\llncssubparagraph\subparagraph
\let\subparagraph\paragraph
\usepackage[compact]{titlesec}
\let\subparagraph\llncssubparagraph
\usepackage{bbm}
\usepackage{mathptmx}
\usepackage{graphicx} 
\usepackage{caption,cite}
\usepackage{subcaption}
\usepackage{mathtools}
\usepackage{amssymb}
\usepackage[square,sort,comma,numbers]{natbib}

\usepackage{tikz}
\usetikzlibrary{arrows}

\newtheorem{definition}{Definition}

\newtheorem{lemma}{Lemma}

\newtheorem{proposition}{Proposition}

\newtheorem{theorem}{Theorem}

\newtheorem{example}{Example}

\newtheorem{remark}{Remark} 

\newtheorem{corollary}{Corollary}

\newcommand*{\QEDB}{\hfill\ensuremath{\square}}%

\ifCLASSINFOpdf
\else
\fi
\begin{document}

\title{A Game-Theoretic Foundation of Deception: Knowledge Acquisition and Fundamental Limits}
%

\author{Tao Zhang and Quanyan Zhu}

\maketitle










\begin{abstract}
Deception is a technique to mislead human or computer systems by manipulating beliefs and information. Successful deception is characterized by the information-asymmetric, dynamic, and strategic behaviors of the deceiver and the deceivee. This paper proposes a game-theoretic framework of a deception game to model the strategic behaviors of the deceiver and deceivee and construct strategies for both attacks and defenses over a continuous one-dimensional information space.
We use the signaling game model to capture the information-asymmetric, dynamic, and strategic behaviors of deceptions by modeling the deceiver as a privately-informed player called sender and the deceivee as an uninformed player called receiver.
We characterize perfect Bayesian Nash equilibrium (PBNE) solution of the game and study the deceivability. We highlight the condition of deceivee's knowledge enhancement through evidences to maintain the equilibrium and analyze the impacts of direct deception costs and players' conflict of interest on the deceivability.

\end{abstract}

\IEEEpeerreviewmaketitle

\section{Introduction}

Deception is a technique used to cause animals \cite{cott1940adaptive}, human \cite{vrij2008increasing, gneezy2005deception} or computer systems \cite{bodmer2012reverse} to have false beliefs. 
The purpose of deception is to mislead the deceivees to behave in a way that is usually against their interests but is favorable to the deceiver. 
Deceptions can be viewed as an approach for attacks as well as defense. 
For example, honeyfile is a common deception tool for defense that creates a fake file directory that behaves like a normal file system of an active user to detect intrusions or malicious insiders.
Attackers could create a website with fake promotions to attract target deceivees to reveal their personal data such as credit card information and social security number.

Successful deception fundamentally depends on the information asymmetry between the deceiver and the deceivee.  Deceivees need to obtain information that is indirect and difficult to verify for decision-making. 
Deceivers can take advantage of this by pretending to be a trustworthy information provider.
It is possible to fool, mislead, or confuse the deceivees. But to do so it may cause them to take some unforeseen or unfavorable actions to the deceivers.
Therefore, successful deception also requires the deceivers to have the ability to acquire information, accurately understand the goals of the deceivees, and make the induced actions predictable.

The deceivers strategically manipulate the private information to suit their own self-interests. The manipulated information is then revealed to the deceivees, who, on the other hand, make decisions about the information received. 
It is important for the deceivee to form correct beliefs based on past observations, take into account the potential damage caused by deception, and strategically use the observed information for decision-making.
If the deception is necessary to achieve the deceivers' goal that would cause damages to the deceivees, the deceivees can then get prepared to invest resources in detecting and denying the deceptions as well as recovering the damage.

Deception is costly. Designing and executing deception requires resource and effort. It is possible that the deceptions incur costs that commensurate with the benefits from achieving the goal. 
It is well to be aware that even the slickest deception plan cannot guarantee the desired behavior from the deceivee \cite{bodmer2012reverse}. Thus, the behavior induced in the deceivee could also incur cost to the deceiver. Other costs include ex-post penalties caused by, for example, breaking the law or regulations.
The cost and risk are central to deception that may change the incentive of the deceiver.

Deception has constraints. In order to make it work across time and space, the deceiver must have a clear and consistent understanding about the nature and the environment of the system associated with the deceivee. Such knowledge include the intelligence of the deceivee, the attack surface of the cyber system, and how the authorized parties behave within the system. 
On the deceivees' side, they can take advantage of the constraints to put serious and continuing effort to gather information on any possible deceptions on the protected system.

In this project, we propose a framework of a deception game to model the strategic behaviors of the deceiver and deceivee and construct strategies for both attacks and defenses over a continuous one-dimensional information space.
We use the signaling game model to capture the information-asymmetric, dynamic, and strategic behaviors of deceptions by modeling the deceiver as a privately-informed player called sender and the deceivee as an uninformed player called receiver.
Additionally, the deceivee is allowed to acquire probabilistic evidence about the deception through investigations.

We parameterize the costs of deceptions and model both players as cost-minimizers. The deceivability of the deception game is analyzed by characterizing the perfect Bayesian Nash equilibrium (PBNE).
The equilibrium solution of the game provides guidances that can be used for the mechanism design of a anti-deception system as well as plans for deception as defense approaches.
Our results show that the deception game admits a class of partial-pooling PBNE that identifies the deceivable and the undeceivable regions of a one-dimensional information space.
We highlight the condition of deceivee's knowledge enhancement through evidences to maintain the equilibrium and analyze the impacts of direct deception costs and the players' conflict of interest on the deceivability.

\subsection{Related Work}

Our deception game is related to a class of security games of incomplete information.
For example, Powell in \cite{powell2007allocating} has considered a game between an attacker and a defender, where the defender has private information about the vulnerability of their targets under protection.
Powell models the information asymmetric interactions between players by a signaling game, and finds a pooling equilibrium where the defender chooses to pool, i.e., allocate resources in the same way for all targets of different vulnerabilities, and the attacker cannot know the true level of vulnerability of all targets.
Brown et al. \cite{brown2005two} have studied a zero-sum game between an attacker and a defender in the scenario of ballistic missile positioning. They have introduced the incomplete information to investigate the value of secrecy by restricting the players' access to information.

Previous literature has also considered deception in a variety of scenarios. 
Pawlick et al., \cite{pawlickanalysis} have considered a class of deception for network security and extended the formulation of signaling game \cite{crawford1982strategic} by including a detector that provides probabilistic evidence of deception. They have analyzed the deceivability in pooling and partially-separating equilibria of the game.
Zhang et al., \cite{zhang2017strategic} have proposed an equilibrium approach to analyze the GPS spoofing in a model of signaling game with continuous type space.
They have found a PBNE with pooling in low types and separating in high types, and provided an equilibrium analysis of spoofing.
The model proposed in Ettinger et al. \cite{ettinger2010theory} have used an equilibrium approach to belief deception in bargaining problems when the agents only have coarse information about their opponent's strategy. 

This work provides a signaling-game theoretic holistic quantitative framework to analyze the interactions between the deceiver and the deceivee and to design mechanisms that improve the cost efficiency for either party.

The rest of the paper proceeds as follows. Section \ref{model} describes our game model and define the concept of deceivability. In Section \ref{equilibrium_sec} and \ref{analysis_deceivability}, we describe the equilibrium concept and studies the deceivability in the equilibrium, respectively. Section \ref{numerical} illustrates the theoretical analysis with numerical examples. Finally, we conclude our work in Section \ref{conclusion}.

\section{Model}\label{model}

Our deception game contains two players: a deceiver (sender, $S$, a.k.a. he) and a deceivee (receiver, $R$, a.k.a. she). 
%
%
The game is information asymmetric because the deceiver privately possesses the target information that is unknown to the deceivee.
There exists \textit{conflict of interest} between two players.
The deceiver costly manipulates the information according to the conflict of interest.
The deceivee strategically reacts based on all the information she has.
The existence of deception cost and specific conflict of interest make the signaling game a \textit{nonzero-sum} game even through there is a non-cooperative relationship between the deceiver and the deceivee.
%


\subsection{States, Messages, Strategies, Actions, and Beliefs}

We consider a game where the informed $S$ communicates his one-dimensional private information to an uninformed $R$. Table \ref{table_1} summarizes the notations.

\begin{table}
\centering
\caption{Summary of notation}\label{table_1}
  \begin{tabular}{ | c | c  |}
    \hline 
    Notation & Meaning \\ \hline \hline 
    $S$, $R$ & Deceiver and Deceivee  \\ \hline
   
    $F(\tilde{\theta})$, $f(\tilde{\theta})$  & distribution, density of $\theta$ \\ \hline
    $\theta\in \Theta$, $m\in M$, $r\in \Theta$ & Signal, Message, Report\\  \hline
       $\Omega(m): M \rightarrow \Theta$ & Reporting Function  \\ \hline
          $\sigma^S(\theta): \Theta \rightarrow \Theta$ & Strategy of $S$  \\ \hline
             $\eta^S(r): \Theta \rightarrow M$ & Message Strategy of $S$  \\ \hline
             $a\in A $ & Action of $R$  \\  \hline 
             
     $\sigma^R(r): \Theta \rightarrow A$ & Strategy of $R$  \\ \hline
     
     $q^S(m|\theta)$ & Conditional distribution induced by $\sigma^S$  \\  \hline
     $\mu^R(\theta|m) $ & (Posterior) Belief of $R$  \\  \hline
     $C^A $ & Cost of $S$ induced by Action $a$  \\  \hline
     $C^D$ & Deception Cost of $S$ \\  \hline
     $C^S = C^A+C^D$ & Total Cost of $S$  \\  \hline
     $C^R$ & Cost of $R$  \\  \hline 
     $e\in E$ & Evidence Acquired by $R$  \\  \hline
     $\gamma(e| \Psi^i, m)$ & Probability of $e$ given event $\Psi^i$ and $m$  \\  \hline
  \end{tabular}
\end{table}

%



\textbf{States.} We use the notion \textit{state} to represent the information that is unknown to $R$. 
$S$ privately possesses the state $\theta \in \Theta \equiv [\underline{\theta},\overline{\theta}]$, with $-\infty< \underline{\theta} < \overline{\theta}< + \infty$. We assume that the state $\tilde{\theta}$ is continuously distributed according to a differentiable probability distribution $F(\tilde{\theta})$, with strictly positive density $f(\theta)$ for all $\theta \in \Theta$. All aspects of the game except the value of the true state $\theta$ are common knowledge.

\textbf{Message and Report.} By the notion \textit{message}, we describe the format of information about the state $S$ communicates to $R$; by the notion \textit{report}, we represent the \textit{value of state} carried by the message.
After privately observing the state $\theta$, $S$ first determines a report $r\in \Theta$ for the true state $\theta$, and then sends $R$ a message $m\in M$, where $M$ is a Borel space of messages.
Let $\Omega: M \rightarrow \Theta$ denote the \textit{report interpretation function} such that $\Omega(m)$ gives the report $r$ carried in $m$.
%
%
%
Given the true state $\theta$, we say $m$ tells the truth if $\Omega(m) = \theta$. 
%
%
We assume that for each state $\theta\in \Theta$, there is a sufficiently large number of messages that yields the same report, and each $m\in M$ has a unique value of report $\Omega(m)$.
In other words, the message space can be partitioned as $M = \cup_{r} M_{r}$, with $|M_{r}| \rightarrow \infty$ for all $r$ and $M_{r} \cap M_{r'} = \emptyset$ if $r\neq r'$, and $\forall m\in M_{r}$, $\Omega(m) = r$. 
This assumption can capture the feature of \textit{rich language} in practical deceptions.
We further assume that message $m$ is formed by \textit{``common language''} that can be understood precisely by both $S$ and $R$. In other words, function $\Omega$ is commonly known by both players.


\textbf{Strategies and actions.} 
Let $\sigma^S: \Theta \rightarrow \Theta$ be the strategy of $S$ such that $r = \sigma^S(\theta)$ determines the report $r$ of the true state $\theta$.
Let $\eta^S: \Theta \times \Theta \rightarrow M$ be the \textit{message strategy} of $S$ associated with $\sigma^S$ such that $m = \eta^S(r)$ selects the message $m$ from $M_{r}$ when the strategy $\sigma^S(\theta)$ determines the report $r$ and the true state is $\theta$.
Given $\theta$, the strategy $\sigma^S(\theta)$ determines the set of messages $M_{\sigma^S(\theta)}$ for $\eta^S$ to choose from, and $\eta^S$ determines which specific message $m\in M_{\sigma^S(\theta)}$ to send.
We assume $\sigma^S(\theta)$ associated with $\eta^S$ induces a conditional probability $q^S(m | \theta)$.
%
After receiving $m$, $R$ chooses an action $a\in A \equiv \Theta$ according to a strategy $\sigma^R: \Theta \times M \rightarrow A$ using $r = \Omega(m)$.
$\sigma^R(r,m)$ gives the action $R$ acts upon the message $m$ (and thus $r=\Omega(m)$). The action $a$ is the final decision of $R$ that represents the inference about the true state.
%
%

%

\textbf{Beliefs.} 
Based on $m$ (and thus $r=\Omega(m)$) and her \textit{prior} belief $f(\theta)$, $R$ forms a \textit{posterior} belief $\mu^R: \Theta \rightarrow [0,1]$ of the true state $\theta \in \Theta$. The posterior belief $\mu^R(\theta|m)$ gives the likelihood with which $R$ believes that the true state is $\theta$ based on $m$. $R$ then determines which action to choose based on her belief $\mu^R$.

 \begin{figure}
 \includegraphics[width=0.5\textwidth]{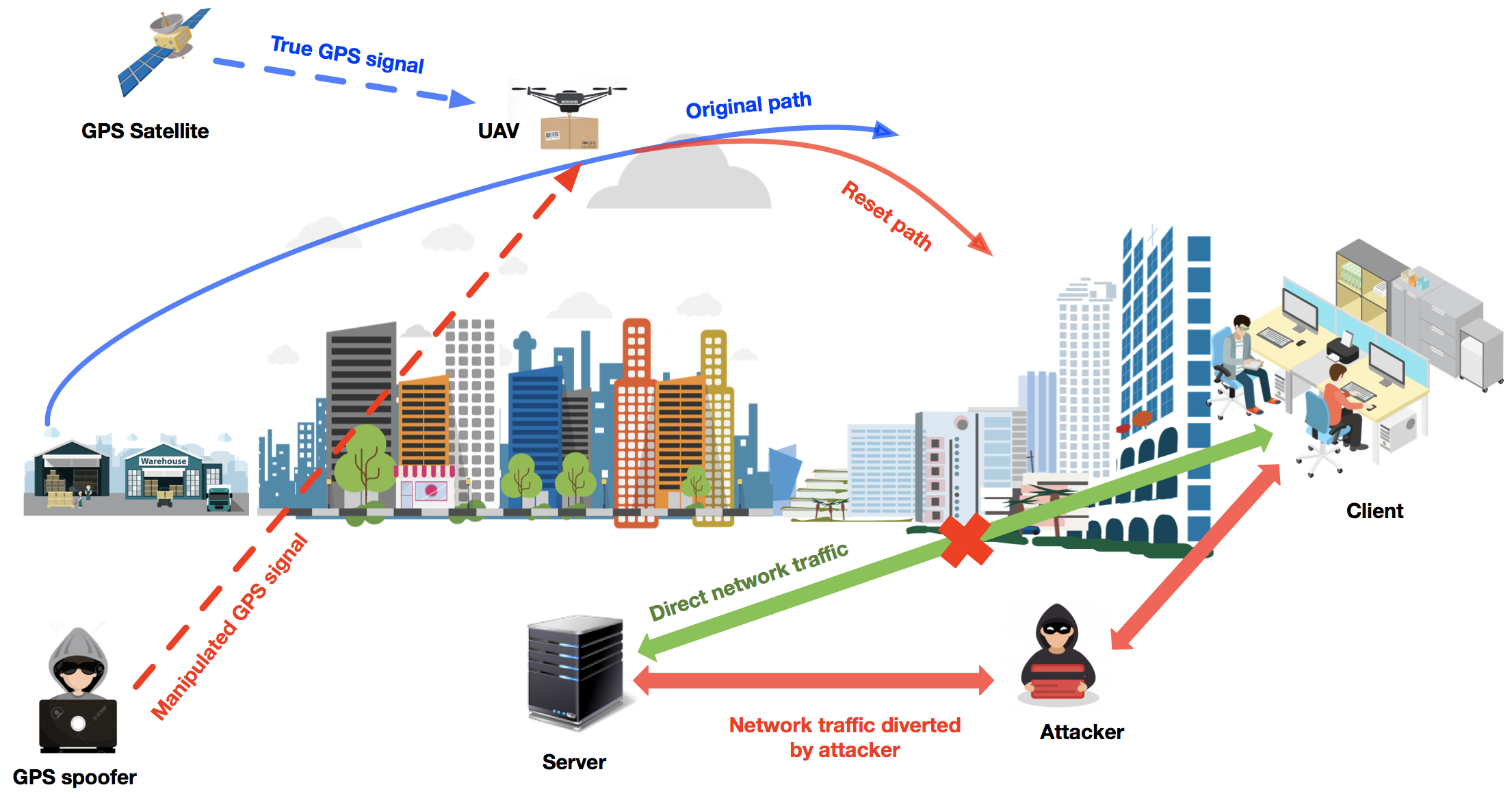}
   \caption{Deception examples: GPS spoofing (left) and man-in-the-middle attack (right).}\label{deception_example}
  \centering
\end{figure}

Potential application scenarios include but are not limited to the following ones.

\begin{example}(\textit{GPS signal spoofing}.)
With reference to Fig. \ref{deception_example}, consider that a GPS signal spoofer launches an spoofing attack targeting on an unmanned aerial vehicle (UAV), whose navigation is solely based on GPS system. Here, the spoofer is the deceiver and the UAV is the deceivee. The latitude (resp. longitude) is the state $\theta^{A}\in \Theta^A$ (resp. $\theta^{O}\in \Theta^O$). The state spaces $\Theta^A$ and $\Theta^O$ identify the activity area of the UAV. The spoofer sends the UAV a manipulated \textit{navigation message} $m$, which contains data including ephemeris parameters, time parameters, and service parameters with satellite health information. 
After privately possessing the true location coordinates $(\theta^A, \theta^O)$ of the UAV, the spoofer's strategy $\sigma^S(\theta^X)$ determines the reported coordinate $r^X$ and the associated message strategy $\eta^S(r^X)$ chooses the navigation message $m$ such that $\Omega(m) = r^X$, for $X=A$, $O$. 
After receiving $m$, the UAV takes an action $a^X$ according to the strategy $\sigma^R(r^X,m)$ for each coordinate such that $a^X$ is the coordinate the UAV chooses to believe, for $X=A$, $O$.
\end{example}

\begin{example}(\textit{Man-in-the-middle attack (MITM)}.)
Suppose an attacker intrudes into the communication between a client and a server as shown in Fig. \ref{deception_example}. He attempts to intercept the data and inject false information. Here the attacker is the deceiver, and the client/server is the deceivee.
The state $\theta\in \Theta$ is the original data, and report $r\in \Theta$ is the false data. The message $m \in M$ is the data displayed in the client's device
%
After privately observing the true data $\theta$ from the server (resp. client), the attacker determines a report $r= \sigma^S(\theta)$, and sends a message $m=\eta^S(r)$ to the client (resp. server).
After receiving $m$, the client (resp. server) takes an action $a\in A\equiv \Theta$ to infer the content of the data.
\end{example}

\subsection{Cost Functions}

Deception is costly endogenously and exogenously. The endogenous cost of deception is caused by the induced action $a$ in $R$, which is given by $C^A(a, \theta; b): A\times \Theta \rightarrow \mathbb{R}$, where $b\in \mathbb{R}$ is the \textit{conflict of interest} between $S$ and $R$ that determines the most desired behavior from $R$ for $S$, i.e., the goal of deception.
$b=0$ features perfect common interest between two players.
We allow a quite general specification of $C^A$, requiring it to only satisfy: for any state $\theta$, $C^A$ is increasing the further the action $a$ is from the most desired action characterized by $b$. Since $b$ is not a decision variable, $C^A(a, \theta; b)$ will be written as $C^A(a, \theta)$ hereafter.
To simplify the analysis, we set $C^A (a, \theta) \equiv (a- (\theta + b))^2$, where $\theta+b$ specifies the  goal of deception. 

The exogenous cost of deception is caused by misrepresenting the true state $\theta$, which takes into account the efforts and resources to generate the report $r$ and the message $m$.
We endogenize the cost directly induced by $m$ in the cost of reporting $r$ by defining the exogenous cost as $C^D(r, \theta):\Theta \times \Theta \rightarrow \Theta$.
Similar to $C^A$, the only requirement for $C^D$ is: for any state $\theta$, the larger $r$ deviates from $\theta$, the larger the cost $C^D$ becomes. 
In this paper, we set $C^D \equiv (r- \theta)^2$.
Next, let $C^S: A\times \Theta \times M \rightarrow \mathbb{R}$ denote the (total) cost function of $S$ defined as
\begin{equation}\label{Cost_S}
C^S(a, \theta, r) \equiv C^A + k C^D,
\end{equation}
where $k$ quantifies the intensity of the deception cost $C^D$.  

On the deceivee's side, let $C^R(a, \theta):  A \times \Theta \rightarrow \mathbb{R}$ denote the cost of $R$. 
$C^R$ takes into account the risk induced by $R$'s misinference of the true state $\theta$ via her action $a$.
The only property for the specification of $C^R$ is: the further $a$ is from $\theta$, the larger $C^R$ becomes.
In this paper, we set $C^R \equiv (a - \theta)^2$ denote the cost $R$ suffers when she takes action $a$ and the true state is $\theta$. 
Define, for all $\theta\in \Theta$,
$$
\alpha^S(\theta) \equiv \arg\min_{a} C^S(a,\theta, r),
$$
and
$$
\alpha^R(\theta) \equiv \arg\min_a C^R(a, \theta), 
$$
that is, $\alpha^R(\theta)$ and $\alpha^S(\theta)$ are two actions taken by $R$ as functions of $\theta$ that are the most preferred by $R$ and $S$, respectively.
Clearly, $\alpha^R(\theta) \equiv \theta$ and $\alpha^S(\theta) \equiv \theta + b$.
For convention, if $C(x_1, x_2)$ is a twice continuously differentiable function, then $C_i(x_1,x_2) = \frac{\partial C}{\partial x_i}$ and $C_{ij} = \frac{\partial^2 C}{\partial x_i \partial x_j}$, for $i,j=1$, $2$.


\subsection{Monotone Deception and Deceivability}

As a key component of deception, misrepresented state is costly generated and processed for a purpose that can be characterized by the conflict of interest $b$. Merely hiding the true states by falsified states can induce unforeseen or unfavorable actions from the deceivee that can increase the endogenous cost.
In this paper, we consider a fixed $b>0$ for all $\theta\in \Theta$. In this case, the deceiver's most preferred action is $\alpha^S(\theta)>\theta$, for all $\theta\in \Theta$, which allows us to restrict attention to a class of monotone \textit{inflated} deception, in which the strategy profile $(\sigma^S, \sigma^R)$ satisfies conditions in the following definition.
\begin{definition}\label{monotone_deception}
A deception with $S$'s strategy $\sigma^S$ and $R$'s belief $\mu^R$ is monotone if
\begin{itemize}
\item $\sigma^S(\theta)$ is a non-decreasing function of $\theta$;
\item $\sigma^R(r, m)$ is a non-decreasing function of $r$.
\end{itemize}
\end{definition}
Definition \ref{monotone_deception} features a class of strategy 
The monotone deception in Definition. \ref{monotone_deception}

The deceivability of the deception are described in the following precise sense.
\begin{definition}
Given the state $\theta \in [\theta^{''}, \theta^{'}]$, $S$'s strategy $\sigma^S(\theta) = r$, and message strategy $\eta^S(r) = m$,
\begin{itemize}
\item $R$ is \textit{undeceivable} over $[\theta^{''}, \theta^{'}]$ if $\sigma^R(r,m) = \alpha^R(\theta) \equiv \theta$, for all $\theta \in [\theta^{''}, \theta^{,}]$. Here, $\sigma^S(\theta) \neq \sigma^S(\theta')$ for all $\theta\neq \theta' \in [\theta^{''}, \theta^{'}]$, and $\eta^S(r) \in M_{r}$. The corresponding $\mu^R$ is \textit{informative}.
%
%
%
%
The interval $[\theta^{''}, \theta^{'}]$ is called \textit{undeceivable region} (UR).

\item $R$ is \textit{deceivable} over $[\theta^{''}, \theta^{'}]$ if the only knowledge she has is that $\theta$ lies in $[\theta^{''}, \theta^{'}]$. 
$R$ chooses $\sigma^R(r,m) = \hat{a}^R(\theta^{''}, \theta^{'})$, by minimizing the expected cost over $[\theta^{''}, \theta^{'}]$, i.e., 
\begin{equation}\label{uninformative_action}
\hat{a}^R(\theta^{''}, \theta^{'}) \in \arg\min_{\sigma^R\in A} \int^{  \theta^{''}}_{ \theta^{'}  } C^R(\sigma^R, \theta) f(\theta) d \theta.
\end{equation}
Here, $\sigma^S(\theta)$ and $\eta^S(r)$, respectively, choose the same report $r$ and the same message $m$, for all $\theta\in [\theta^{''}, \theta^{'}]$. Thus, given $m$, $q^S(m|\theta)$ is the same for all $\theta\in [\theta^{''}, \theta^{'}]$, where $q\in (0,1)$. The corresponding $\mu^R$ is \textit{uninformative}.
The interval $[\theta^{''}, \theta^{'}]$ is called \textit{deceivable region} (DR). 
\end{itemize}
\QEDB
\end{definition}

\subsection{Knowledge Acquisition: Evidence}\label{sec_evidence}

We allow $R$ to acquire additional knowledge through investigations when the state is in a DR, $[\theta^{''}, \theta^{'}]$, by partitioning it into multiple intervals, denoted by a strictly increasing sequence, $<\theta_0 =\theta^{''}, \theta_1, \dots, \theta_J =  \theta^{'}>$. Then $R$ conducts investigations for each interval. 
In this paper, we consider the case when there are two investigation intervals in order to simplify the analysis. 

\begin{figure}
 \includegraphics[width=0.5\textwidth]{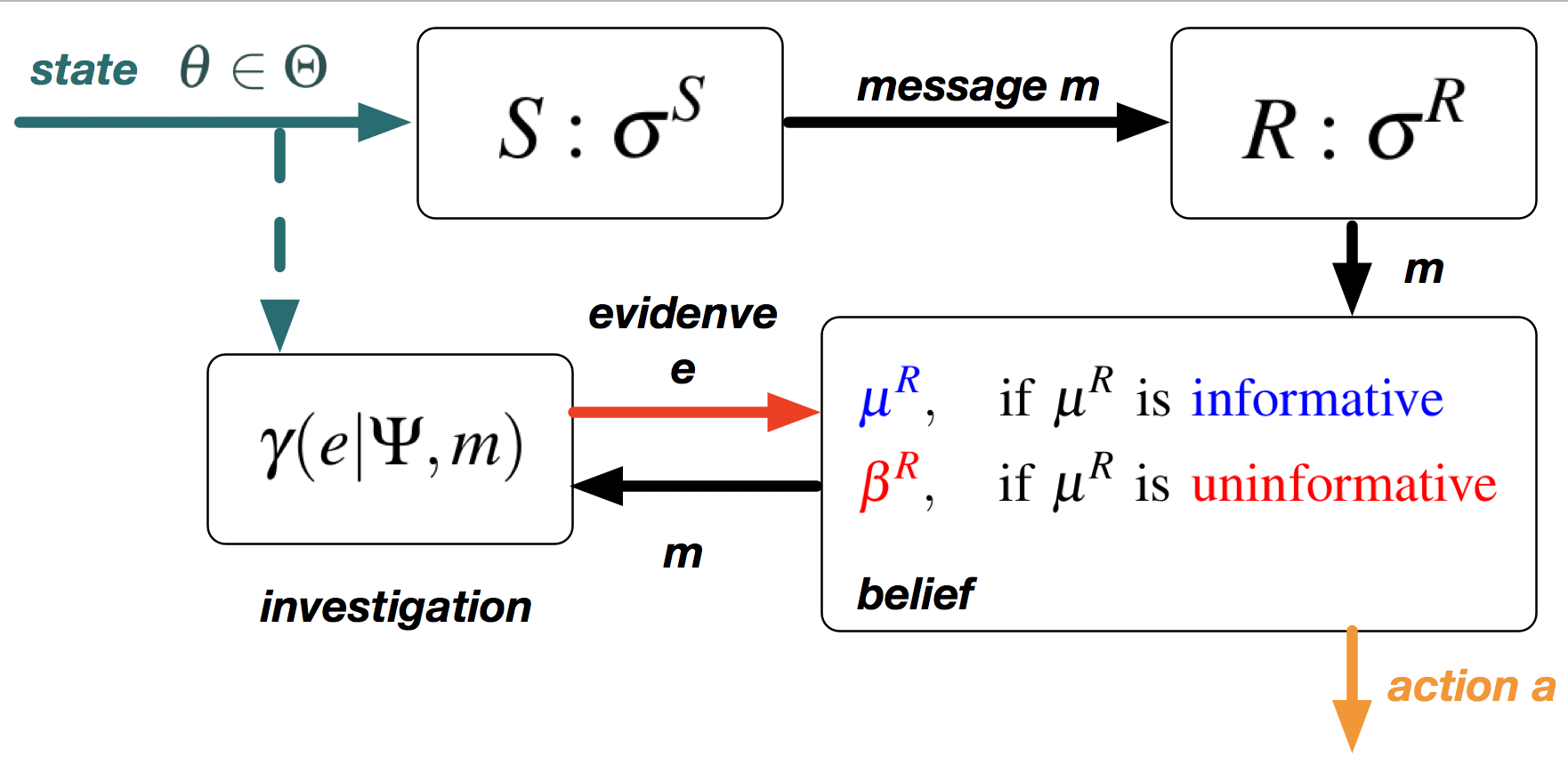}
   \caption{Signaling games with evidence acquisition by investigation. The probability $\gamma(e| \Psi, m)$ of emitting evidence $e$ depends on the event $\Psi$ and the message $m$ sent by $S$. If the belief $\mu^R$ is informative, $\mu^R$ is used; if $\mu^R$ is uninformative, $\beta^R$ is used as the posterior.}\label{SG_block}
  \centering
\end{figure}

Let $\sigma^c \in (\theta^{''}, \theta^{'})$ be the \textit{investigation partition state} such that $[\theta^{''}, \theta^{'}]$ is partitioned into two non-overlapping investigation regions $\Theta^0 = [\theta^{''}, \theta^c]$ and $\Theta^1 = [\theta^c, \theta^{'}]$.
Let $\Psi \in \Gamma = \{\Psi^0, \Psi^1\}$, where $\Psi^i$ denote the event $\{\theta\in \Theta^i\}$, for $i=0$, $1$, with the probability $P(\Psi^i) = \frac{ \int_{\Theta^i} f(\tilde{\theta}) d \tilde{\theta} }{ \int_{[\theta^{''}, \theta^{'}]} f(\tilde{\theta})d \tilde{\theta}}$.
The investigation for $\Theta^0$ and $\Theta^1$ generates noisy evidence $e\in E = \{0,1\}$, where $e=i$ represents $\Psi^i$, for $i=0$, $1$. Suppose that the investigation emits evidence by the probability $\gamma(e|\Psi, m)$.
Let $x = \gamma(e=0| \Psi^0, m) $ and $y = \gamma(e = 1 | \Psi^1 ,m)$ be the two true positive rates, which are private information of $R$. 
%
With a slight abuse of notation, let $\sigma^R(\Psi, m, e):\Gamma \times M \times E \rightarrow A$ be the strategy of $R$ with evidence $e$.
Fig. \ref{SG_block} depicts the signaling game model for the deception with knowledge acquisition through investigation.
%


\section{Equilibrium}\label{equilibrium_sec}

In game theory, the concept of equilibrium defines a strategy profile in which each player best responds to the optimal strategy of other player(s).
In any equilibrium, no player has incentive to deviate from the equilibrium strategy.
We consider the perfect Bayesian Nash equilibrium (PBNE) \cite{osborne1994course} as our solution concept.
PBNE captures the information asymmetry between two players and asynchronous optimizations of the players. 
Definition \ref{PBNE_1} defines the PBNE for the deception game with evidence.
\begin{definition}\label{PBNE_1}
(Perfect Bayesian Nash Equilibrium) A PBNE of the game is a strategy profile $(\sigma^S, \sigma^R)$ and a posterior belief system $(\mu^R,\beta^R)$ that satisfy the following conditions:
\begin{itemize}
\item \textit{(Deceiver's Sequential Rationality)} $S$ minimizes his expected cost given the deceivee's strategy $\sigma^R$ and the distribution of the evidence $e$: for each $\theta\in \Theta$,
\begin{equation}\label{SR_S}
\sigma^{S*}( \theta) \in \arg\min_{\sigma^S} C^S(\sigma^{R*}, \theta, \sigma^S ).
\end{equation}

\item \textit{(Deceivee's Sequential Rationality)} $R$ minimizes her expected cost given $S$'s strategy $\sigma^{S*}$ and her posterior belief $\mu^R(\theta| m)$: for any $m \in M$,
\begin{itemize}
\item if $\mu^R(\theta| m)$ is informative, i.e., Eq. (\ref{CB_R}),
{\small
\begin{equation}\label{SR_R}
\sigma^{R*}(r, m) \in \arg\min_{\sigma^R\in A} \int_{\theta\in\Theta} C^R(\sigma^R, \theta) \mu^R(\theta|m)d \theta;
\end{equation}
}
%
%
\item if $\mu^R(\theta| m)$ is uninformative over $\Theta_U \equiv [\theta^{''}, \theta^{'}] \subseteq \Theta$, i.e., Eq. (\ref{uninformative_mu}), 
{\small
\begin{equation}\label{SR_R_evidence}
\sigma^{R*}(\Psi, m,e) \in \arg\min_{\hat{a}^i}  \sum_{i=0}^1 \int_{\Theta^i_U } \beta(\Psi^i| m, e) C^R(\hat{a}^i, \tilde{\theta})f(\tilde{\theta})  d\tilde{\theta},
\end{equation}
}

\noindent where $\theta^0_U\equiv [\theta^{''}, \theta^c]$, $\theta^1_U \equiv [\theta^{''}, \theta^c]$, and $\hat{a}^i \equiv \arg\min \int_{\theta^i_U}C^R(a, \tilde{\theta})d\tilde{\theta}$. 
\end{itemize}

%


\item \textit{(Consistent Belief)} The posterior belief of $R$ is updated according to Bayes' rule, as
\begin{equation}\label{CB_R}
\mu^R(\theta | m ) = \frac{f(\theta) q^S(m|\theta)}{\int_{\Theta} f(\tilde{\theta}) q^S(m|\tilde{\theta})d \tilde{\theta}}.
\end{equation}
If $\int_{\Theta} f(\tilde{\theta}) q^S(m|\tilde{\theta})d \tilde{\theta}= 0$, $\mu^R(\theta|m)$ may be set to any probability distribution over $\Theta$.

If $\mu^R$ is \textit{uninformative}, i.e., 
\begin{equation}\label{uninformative_mu}
\mu^R = \frac{f(\theta)}{\int_{[\theta^{''}, \theta^{'}]}f(\tilde{\theta}) d \tilde{\theta}},
\end{equation}
$R$ acquires evidence through investigation, and updates belief using evidence as,
\begin{equation}\label{belief_evidence}
\beta^R(\Psi | e, m) = \frac{\gamma(e|\Psi, m) P(\Psi)}{\sum_{j=0}^1 \gamma(e|\Psi^j, m) P(\Psi^j) },
\end{equation}
and if $\sum_{j=0}^1 \gamma(e|\Psi^j, m) P(\Psi^j) =0$, $\beta^R(\Psi | e, m)$ may be set to any probability distribution over $\Theta$.

%

%
%
\end{itemize}
\QEDB
\end{definition}

There are three classes of equilibria: separating equilibrium (S-PBNE), pooling equilibrium (P-PBNE), and partial-pooling equilibrium (PP-PBNE). In separating equilibrium, the deceiver sends message $m$ with different values of report $\Omega(m)$ for different states. Separating equilibria are also called \textit{revealing} equilibria because the strategic deceivee can infer the true state even if $\Omega(m)$ does not tell the truth.
In pooling equilibrium, the deceiver sends message $m\in M_{r}$ with the same value of report $\Omega(m)=r$ for all states. 
In partial-pooling equilibrium, however, the deceiver sends the message with the same report for some states and different reports for other states. 
Clearly, the PBNE strategy $\sigma^{S*}$ associated with 
a DR (resp. UR) is pooling (resp. separating) strategy. 

\section{Analysis of Deceivability at Equilibria}\label{analysis_deceivability}

We present the analysis in three steps. In Subsection \ref{incnetives}, we identify the conditions about whether $S$ has incentive to deceive $R$ in equilibria.
In Subsection \ref{sec_UR}, we find the undeceivable region with the property of the corresponding cost-minimizing strategy of $S$. 
In Subsection \ref{sec_PUR}, we find the partially deceivable region with the necessary boundary conditions.

\subsection{Incentives of Deception}\label{incnetives}
In order to deceive $R$ over the DR $[\theta^{'}, \theta^{''}]$, $S$ plays P-PBNE strategy $\sigma^{S*}(\theta) = r$, for all $\theta$ in the DR, and the associated $\eta^S(\theta)$ chooses the same $m$ for all $\theta$ in the DR. 
This P-PBNE strategy induces the action $\hat{a}^R= \sigma^{R*}(\Psi, m,e)$ given in Eq. (\ref{SR_R_evidence}). However, if $S$ uses S-PBNE strategy for $[\theta^{'}, \theta^{''}]$, $R$ plays $a^* = \theta$.
%
%
%
Since both players are cost-minimizers, the incentive of $S$ to deceive $R$ depends on his cost given $R$'s equilibrium action. If the cost induced by $\hat{a}^R$ is more costly than that induced by $a^{*}$, then $S$ has no incentive to deceive $R$, and vice versa. Definition \ref{NITD} summarizes the property of \textit{no incentive to deceive}.
\begin{definition}\label{NITD}
The equilibrium strategy profile $(\sigma^{S*}, \sigma^{R*})$ satisfies the property of \textit{No incentive to deceive} (NITD) over $[ \theta^{'}, \theta^{''} ]$, 
$$
 C^A(\hat{a}^R, \theta) > C^A(\theta, \theta),
$$
for all $\theta\in  [\theta^{'}, \theta^{''}] \subseteq \Theta$, where $\hat{a}^R$ is given by the strategy $\sigma^{R*}(\Psi^i, m,e)$ in Eq. (\ref{SR_R_evidence}).
%
\QEDB
\end{definition}

\subsection{Undeceivable Region}\label{sec_UR}


In this subsection, we consider if there exists any PBNE UR in $\Theta = [\underline{\theta}, \overline{\theta}]$. 
From the definition of UR, the equilibrium strategy of $R$ gives the most preferred action, $\alpha^R(\theta)\equiv \theta$, for all $\theta$ in the UR.
Therefore, in any differentiable S-PBNE, the cost $C^S$ and the strategy $\sigma^S$ have to satisfy the following first-order condition for optimality given $\sigma^{R*}(\theta) = \alpha^R(\theta)$ according to the sequential rationality:
{\small
\begin{equation}\label{first_order}
C^S_1(\alpha^R(\theta), \theta, \sigma^S(\theta))\frac{ d\alpha^R(\theta) }{d \theta} + C^S_3(\theta, \theta, \sigma^S(\theta)) \frac{d \sigma^S(\theta)}{d \theta} = 0.
\end{equation}
}
%
Lemma \ref{lemma_UR_boundry} summarizes the property of the strategy $\sigma^{S*}$ in any UR.

\begin{lemma}\label{lemma_UR_boundry}
If $[\theta_s, \theta_l]$ is an undeceivable region, then for each $\theta \in [\theta_s, \theta_l]$, the equilibrium strategy $\sigma^{S*}(\theta) > \theta$ and it is a unique solution of
\begin{equation}\label{UR_de}
\frac{ d \sigma^S(\theta) }{d \theta} = \frac{ b   }{ k \big(\sigma^S(\theta)  -\theta \big)    },
\end{equation}
with initial condition $\sigma^{S*}(\theta_s)=\theta_s$.
\QEDB
\end{lemma}

\begin{remark}
The proof is in the full version of this paper. Since $\frac{d \sigma^S(\theta)}{d \theta} \geq 0$ and $C^S_1 = -2b$ is independent of $\sigma^S$, there is no strategy $\sigma^S$ such that $C^S_1 \frac{d \alpha^R(\theta)}{d \theta} = 0$ when $C^S_3 = 0$. Then, Eq. (\ref{UR_de}) can be obtained from the first order condition shown in Eq. (\ref{first_order}).
Since $\frac{d \sigma^S(\theta)}{d \theta} \geq 0$, we have $\frac{d \sigma^{S*}(\theta)}{d \theta} > 0$, which means that the optimal strategy of $S$ is strictly increasing.   \QEDB
\end{remark}
Lemma \ref{lemma_UR_boundry} underlies the following proposition.

\begin{proposition}\label{proposition_no_UR}
With initial condition $\sigma^{S*}(\underline{\theta}) = \underline{\theta}$, there exists a cut-off state $\hat{\theta}<\overline{\theta}$ such that a unique solution $\sigma^{S*}$ to Eq. (\ref{UR_de}) is well-defined on $[\underline{\theta}, \hat{\theta}]$ with $\sigma^{S*}(\hat{\theta}) = \overline{\theta}$, and there is no solution to Eq. (\ref{UR_de}) on $(\hat{\theta}, \overline{\theta}]$. 
%
%
\QEDB
\end{proposition}

\begin{remark}
The proof is in the full version of this paper.
Proposition \ref{proposition_no_UR} notes that in S-PBNE, the optimal strategy $\sigma^{S*}$ of $S$ has to choose a report $r$ that is strictly larger than the true state $\theta$, but eventually $\sigma^{S*}$ runs out of such report for $\theta>\hat{\theta}$.\QEDB
\end{remark}

%
 

In any UR at equilibria, $R$'s posterior belief is informative. The endogenous cost induced by the equilibrium strategy of $R$ together with the exogenous deception cost $C^D$ eliminates the incentive of $S$ to deceive $R$ in any UR at equilibria.

\subsection{Deceivable Region}\label{sec_PUR}

Proposition \ref{proposition_no_UR} implies that there is no S-PBNE strategy of $S$ for all $\theta> \hat{\theta}$, because there are not enough states to support the monotone S-PBNE strategy of $S$ for the state in $(\hat{\theta},\overline{\theta}]$.
This suggests a class of PP-PBNE for the state space $\Theta$, which is separating in low states and pooling in higher states.
%
%
%
For convention, let $\sigma^{S,p}: \Theta \rightarrow \Theta$ and $\eta^{S,p}$, respectively, denote the P-PBNE strategy and the associated message strategy of $S$. 
We define this class of PP-PBNE by introducing a boundary state as follows.
\begin{definition}\label{PP-PBNE_SLAPH}
We say that the strategy $\sigma^S$ is a SLAPH (Separating in Low states And Pooling in High states) strategy if there exists a boundary state $\theta_B \in [\underline{\theta}, \hat{\theta}]$ such that 
\begin{itemize}
\item \textit{(S-PBNE)} $\sigma^{S*}(\theta) = r$ with $\eta^{S*}(r)\in M_{r}$, for all $\theta \in [\underline{\theta}, \theta_B)$, and $\sigma^{S*}(\theta) \neq \sigma^{S*}(\theta')$ for all $\theta\neq \theta' \in [\underline{\theta}, \theta_B)$; 
\item \textit{(P-PBNE)} $\sigma^{S*,p}(\theta) = \overline{\theta}$ with $\eta^{S*,p}(\theta) \in M_{\overline{\theta}}$, for all $\theta \in [\theta_B, \overline{\theta}]$.
\end{itemize}
\QEDB
\end{definition}

In any SLAPH equilibrium, both players have no incentive to deviate from the equilibrium strategies.
This requires the boundary state $\theta_B$ to be consistent in the sense that the equilibrium at $\theta_B$ is well-defined. 
Specifically, the cost of $S$ has to satisfy the following \textit{boundary consistency} (BC) condition at $\theta_B$:
\begin{equation}\label{boundary_const}
C^S(\sigma^{R*}(\sigma^{S,p}(\theta_B),m_p), \theta_B, m_p) = C^S(\alpha^R(\theta_B), \theta_B, m_s),
\end{equation}
where $m_p \in M_{\overline{\theta}}$ and $m_s \in M_{\sigma^{S*}(\theta_B)}$.
The BC condition implies that $S$ is indifferent between 
sending $m_p\in M_{\overline{\theta}}$ with $a^* = \sigma^{R*}(\sigma^{S*}(\theta_B), m_p)$ and sending $m_s \in M_{\sigma^{S*}(\theta_B)}$ with action $a^* = \theta_B$. 

The conflict of interest, $b$, is a cost-relevant parameter for $S$ that can induce incentives for $S$ to reveal partial information about any state $\theta\in [\theta_B, \overline{\theta}]$ to $R$ while his cost-minimizing P-PBNE strategy $\sigma^{S*}$ is maintained.
This can be achieved based on the assumption $|M_{\overline{\theta}}| \rightarrow \infty$ and the fact that $C^D$ is equally expensive for all the messages chosen for all state $\theta\in [\theta_B, \overline{\theta}]$.
Specifically, the P-PBNE region $[\theta_B, \overline{\theta}]$ can be further partitioned into multiple pools.
First, some notations for describing the multiple pools are needed. Let $\Theta^P \equiv \big(\theta_0, \theta_1,\dots, \theta_{K-1}, \theta_K\big)$ be a partition of $[\theta_B, \overline{\theta}]$, with $\theta_0 = \theta_B < \theta_1<\dots< \theta_K = \overline{\theta}$. We call each interval $\Theta_{j,j+1} = [\theta_j, \theta_{j+1}]$ is a \textit{pool}.
With an abuse of notation, let $\eta^{S*,p}(\sigma^{S*}(\theta), \theta)$ denote the message strategy that chooses a message $m \in M_{\sigma^{S*}(\theta)}$ for a state $\theta$. In each pool $\Theta_{j,j+1}$, $\eta^{S*,p}(\overline{\theta}, \theta)$ chooses the same message $m\in M_{\overline{\theta}}$, for all $\theta\in \Theta_{j, j+1}$, $j=0,\dots, K-1$.
%


\begin{figure*}
    \centering
      ~ 
    \begin{subfigure}[b]{0.3\textwidth}
        \includegraphics[width=\textwidth]{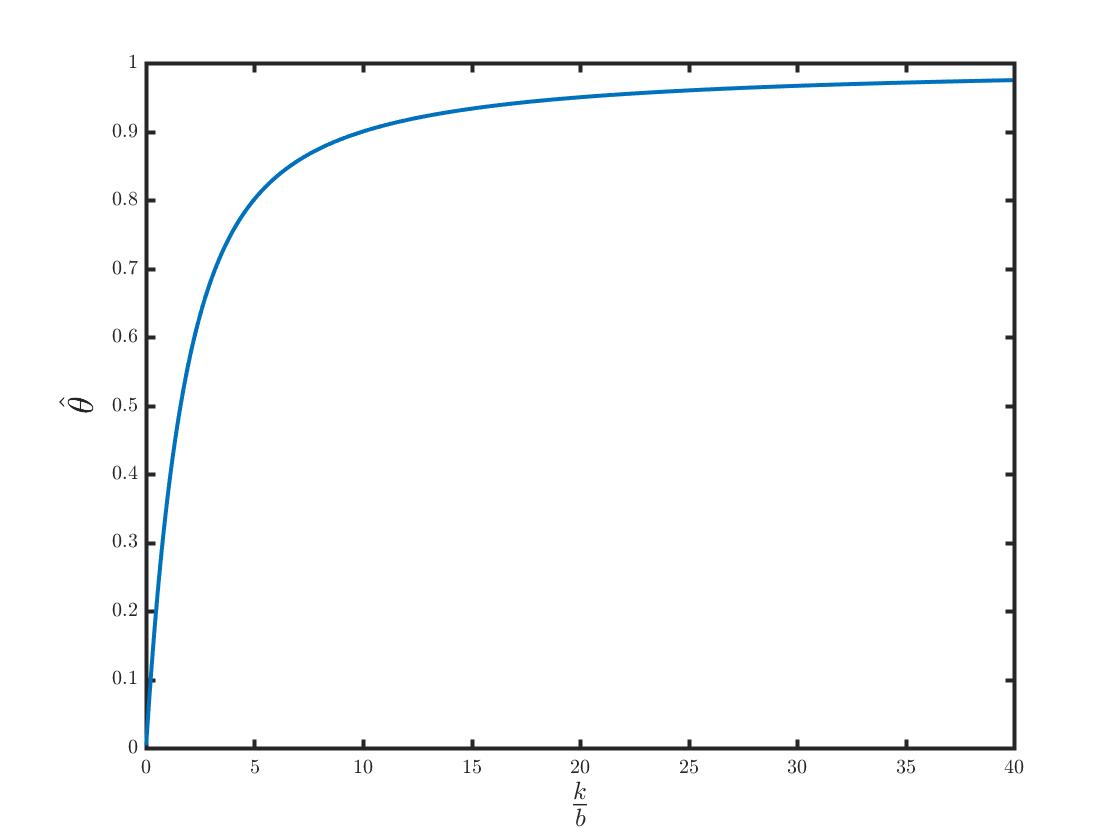}
\caption{$\frac{k}{b}$ vs. $\hat{\theta}$.  $\hat{\theta}$ increases and approaches $1$ as $\frac{k}{b}$ increases.}%
\label{hat_vs_kb}
    \end{subfigure} 
      ~ 
    \begin{subfigure}[b]{0.3\textwidth}
        \includegraphics[width=\textwidth]{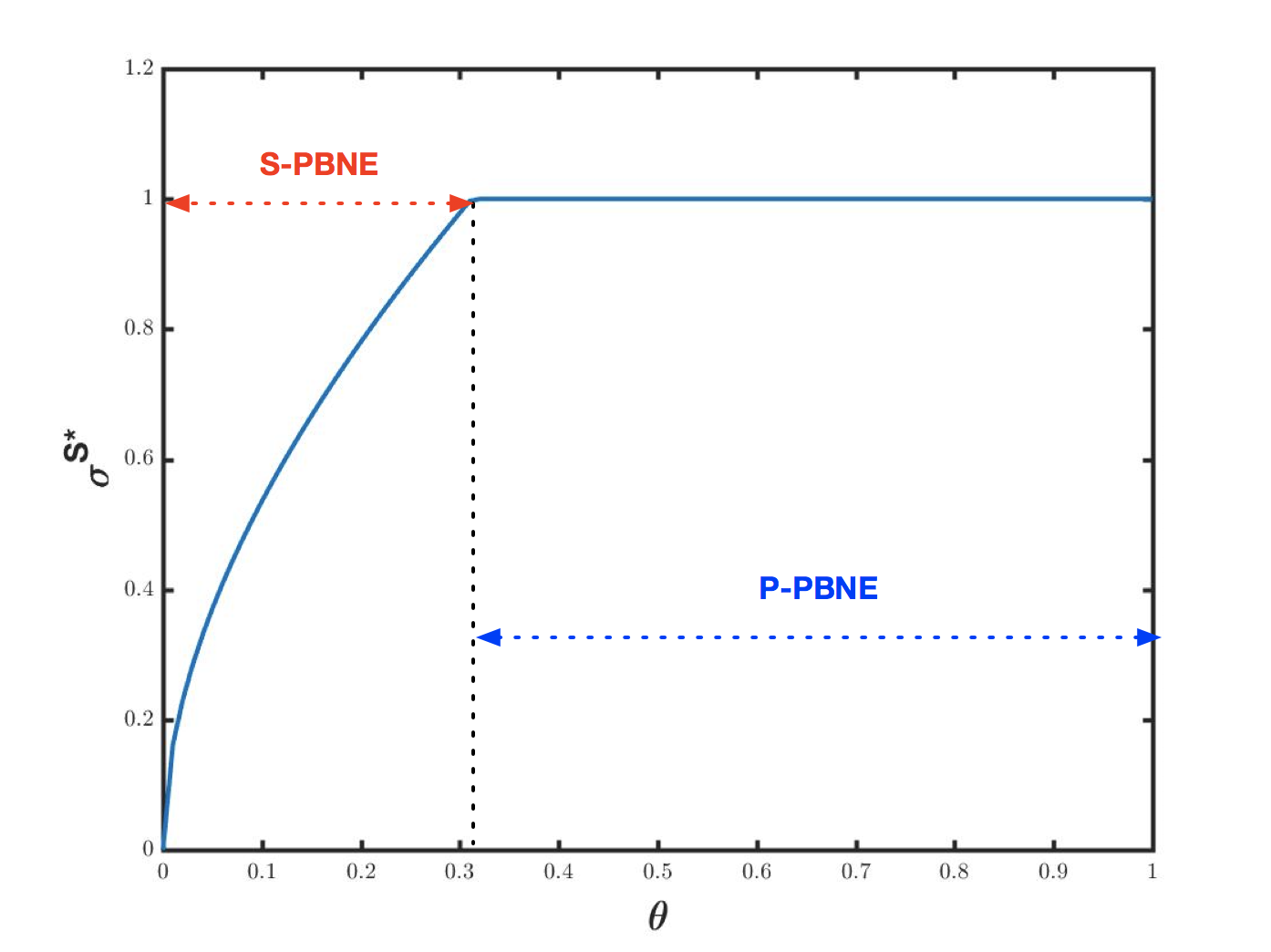}%
\caption{SLAPH. Here, $b=\frac{1}{8}$ and $k=0.1$. The cut-off state is $\hat{\theta} = 0.3117$.}%
\label{SP_PBNE}
    \end{subfigure}
   \begin{subfigure}[b]{0.3\textwidth}
   \vspace{2mm}      \includegraphics[width=\textwidth]{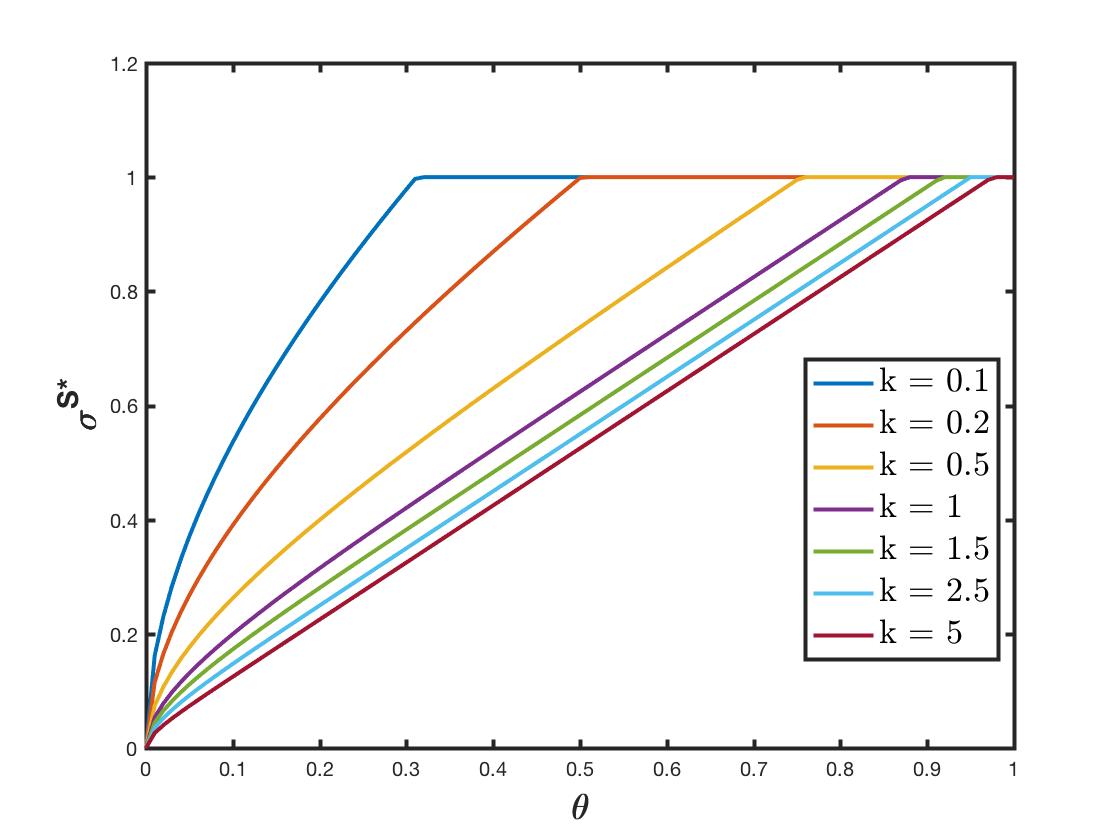}%
\caption{SLAPH. Changes of $\sigma^{S*}$ and $\sigma^{S*, p}$ for $b=\frac{1}{8}$ and different value of $k$.}%
\label{sigma_vs_theta}
    \end{subfigure}
    \caption{}\label{fig_vs}
\end{figure*}
%


Without such multiple pools, the posterior belief $\mu^R(\theta|m)$ shown in Eq. (\ref{CB_R}) can only provide the likelihood of the event $\{\theta \in [\theta_B, \overline{\theta}]\}$.
With multiple pools, on the other hand, $\mu^R(\theta|m)$ can narrow down the location of the true state to a specific pool $\Theta_{j,j+1}$ by yielding the likelihood of the event $\{\theta\in \Theta_{j,j+1}\}$. 
After $R$ determines a pool $\Theta_{j,j+1}$, she acquires evidence $e\in\{e_0, e_1\}$ through investigations by dividing $\Theta_{j,j+1}$ into two sub-intervals $\Theta^0_{j,j+1} \equiv [\theta_j, \theta^I_{j,j+1}]$ and $\Theta^1_{j,j+1} \equiv [\theta^I_{j,j+1}, \theta_{j+1}]$. 
Let $\Psi_{j,j+1}\in \Gamma_{j,j+1}\equiv \{\Psi^0_{j,j+1}, \Psi^1_{j,j+1} \}$ such that $\Psi^i_{j,j+1}$ represents the event $\{\theta\in \Theta^i_{j,j+1}\}$, with probability 
$$
P(\Psi^i_{j,j+1}) = \frac{\int_{\Theta^i_{j,j+1}} f(\tilde{\theta}) d\tilde{\theta} }{  \int_{\Theta_{j,j+1}} f(\tilde{\theta}) d\tilde{\theta}      },
$$
for $i=0$, $1$. 
%
%
On the equilibrium path, $R$ must play $\sigma^{R*}(\Psi_{j,j+1}, m_j, e)$ as defined in Eq. (\ref{SR_R_evidence}) for any $m_{j}$ such that $\eta^{S*,p}(\overline{\theta}, \theta)=m_{j}$ for all $\theta \in \Theta_{j,j+1}$. 
Define 
\begin{equation}
\hat{a}^i(\theta_j,\theta_{j+1}) \equiv \arg\min_{a}\int_{\Theta^i_{j,j+1}}C^R(a, \theta)f(\theta)d\theta, 
\end{equation}
for $i=0$, $1$.
For brevity, define the following notation:
{\small
\begin{equation}\label{cost_efficient_R}
\begin{aligned}
\Delta_{j,j+1} & \equiv \\
&\sum_{e\in E}\sum_{i=0}^1 \int_{\Theta^i_{j,j+1} } \beta(\Psi^i_{j,j+1}| m_j,e) C^R(\sigma^{R*}(\Psi^i_{j,j+1}, m_j, e), \tilde{\theta})  d\tilde{\theta}  \\
- &\int_{\Theta_{j,j+1}} C^R(\hat{a}^R(\theta_j, \theta_{j+1}), \theta) d\theta,
\end{aligned}
\end{equation}
}
\begin{equation}
\delta^0_{j,j+1} = \int_{\Theta^0_{j,j+1}}\big( C^R(\hat{a}^0_{j,j+1}, \theta) - C^R(\hat{a}^1_{j,j+1}, \theta) \big)f(\theta) d\theta,
\end{equation}
\begin{equation}
\delta^1_{j,j+1} = \int_{\Theta^1_{j,j+1}}\big( C^R(\hat{a}^1_{j,j+1}, \theta) - C^R(\hat{a}^0_{j,j+1}, \theta) \big)f(\theta) d\theta.
\end{equation}

\noindent 
Here, $\Delta_{j,j+1}$ gives the expected gain to $R$ over the DR $[\theta_j, \theta_{j+1}]$ by using evidence $e$, $\delta^i_{j,j+1}$ gives the expected benefit to $R$ for correctly inferring $\Psi^i_{j,j+1}$, for $i=0$, $1$.
We say that the investigation is \textit{cost efficient} if the $\Delta_{j,j+1}\geq 0$, for all $j=0,\dots, K-1$, where $\hat{a}^R(\theta_j, \theta_{j+1})$ is defined in Eq. (\ref{uninformative_action}). Otherwise, $R$ has no incentives to acquire the evidence $e$. 
Furthermore, we say that the investigation is \textit{reliable} if $\delta^i_{j, j+1}\geq 0$, for all $j=0,\dots, K-1$.

The necessary and sufficient conditions for the existence of SLAPH equilibrium are summarized in the following theorem.

\begin{theorem}\label{theorem_SLAPH} 
(\textbf{Necessary condition.}) In any SLAPH equilibrium, there exists a boundary state $\theta_B$ such that the pooling interval $[\theta_B, \overline{\theta}]$ can be partitioned into multiple pools denoted by a strictly increasing sequence
$\big(\theta_0, \theta_1,\dots, \theta_{K-1}, \theta_K\big)$ with $\theta_0 = \theta_B$ and $\theta_K = \overline{\theta}$, 
%
%
such that, for all $j=0,\dots, K-1$,  
{\small
\begin{equation}\label{theorem_connect}
C^A(\bar{a}(\theta_j, \theta_{j+1}), \theta_{j+1}) = C^A(\bar{a}(\theta_{j+1}, \theta_{j+2}), \theta_{j+1}),
\end{equation}
}
\begin{equation}\label{theorem_boundary}
C^S(\bar{a}(\theta_0, \theta_1), \theta_B, \overline{\theta}) = C^S(\theta_B, \theta_B, \sigma^{S*}(\theta_B)), \;\; \text{ if } \theta_B>\overline{\theta},
\end{equation}
where $\bar{a}(\theta_j,\theta_{j+1}) = \sum_{i=0}^1P(\Psi^i)\hat{a}^i(\theta_j,\theta_{j+1})$, for all $j=0,\dots, K-1$.

(\textbf{Sufficient Condition.}) Given the multiple-pool PBNE characterized by 
Eq. (\ref{theorem_connect})-(\ref{theorem_boundary}), and if $\theta_B = \underline{\theta}$ and
\begin{equation}\label{sufficient_cond}
C^S(\bar{a}(\theta_0, \theta_1), \underline{\theta} , \overline{\theta}) \geq C^S(\alpha^R(\underline{\theta}),\underline{\theta}, \sigma^{S*}(\underline{\theta})),
\end{equation}
there exists a SLAPH equilibrium.
\QEDB
\end{theorem}


\begin{figure*}\centering
  \centering
    \includegraphics[width=0.9\textwidth]{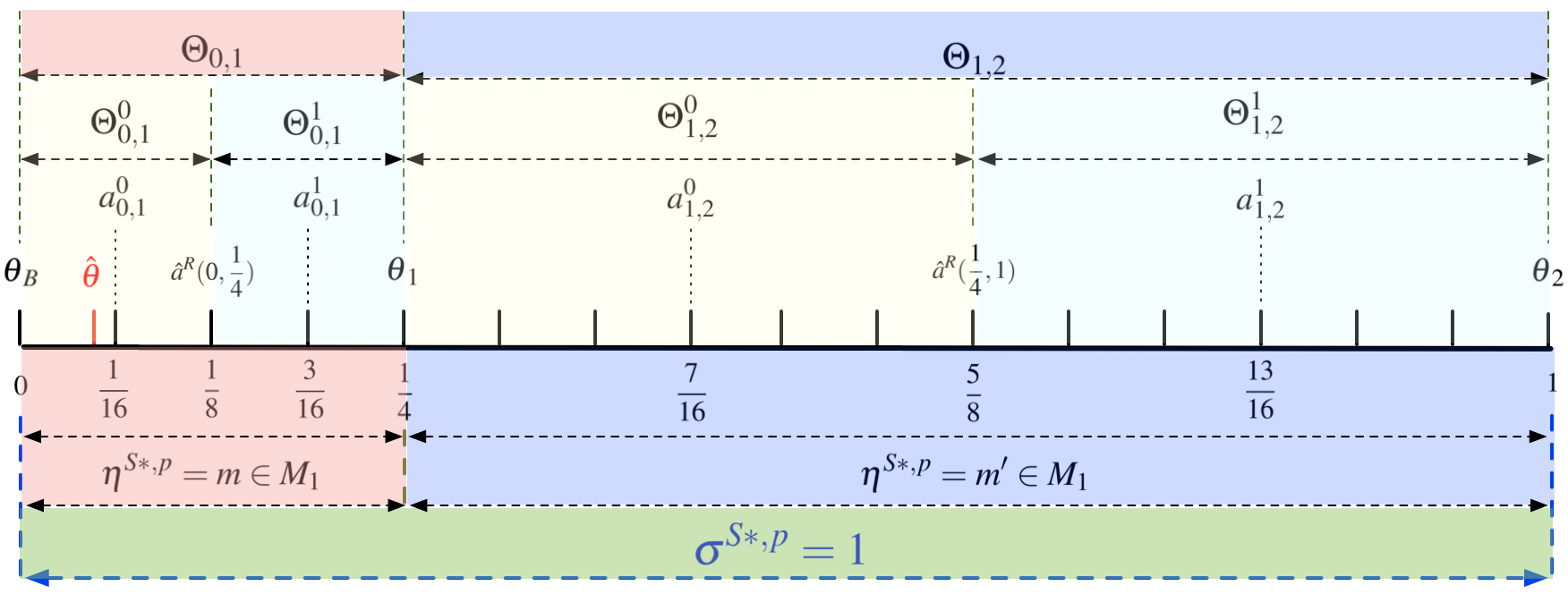}
     \caption{Example of multiple-pool P-PBNE. Here, $\hat{\theta}=\frac{1}{4}$, $\theta_B=0$. There are two pools, i.e., $\Theta_{0,1}=[0,\frac{1}{4}]$ and $\Theta_{1,2} = [\frac{1}{4}, 1]$. $S$ plays $\sigma^{S*,p} = 1$ $\forall \theta\in [0,1]$, and $\eta^{S*,p}(1, \theta) = m$ and  $\eta^{S*,p}(1, \theta) = m'$ with $m\neq m' \in M_1$ for all $\theta\in \Theta_{0,1}$ and $\theta\in \Theta_{1, 2}$, respectively.
     Each pool is partitioned into two investigation regions, i.e, $\Theta^0_{i,j}$ and $\Theta^1_{j,k}$, $i\neq j \neq k = \{0,1,2\}$. In this example, the investigation partition state $\theta^c_{0,1} = \hat{a}^R(0, \frac{1}{4}) = \frac{1}{8}$ and $\theta^I_{1,2} = \hat{a}^R(\frac{1}{4}, 1) = \frac{5}{8}$.} \label{example_multi_pool}
\end{figure*}


Note that the deception cost $C^D$ is equally expensive for all $\theta\in [\theta_B, \overline{\theta}]$. 
Eq. (\ref{theorem_connect}) says that at each link state $\theta_{j+1}$ connecting $\Theta_{j,j+1}$ and $\Theta_{j+1, j+2}$, the costs induced by $\bar{a}(\theta_j, \theta_{j+1})$ and $\bar{a}(\theta_{j+1}, \theta_{j+2})$, respectively, should keep the same. Otherwise, $S$ has incentive to deviate from the current partition to combine these two consecutive pools by sending the message that induces cheaper $C^A$ but the same $C^D$. This is not ideal for $R$ because larger pools make the posterior less informative that could increase the cost for $R$.
Similarly, Eq. (\ref{theorem_boundary}) says that at the boundary state $\theta_B$, $S$ should be indifferent between playing S-PBNE strategy and inducing action $\sigma^{R*}(\theta_B)$ versus playing P-PBNE and introducing action $\bar{a}(\theta_0, \theta_1)$.
Inequality (\ref{sufficient_cond}) notes that if the boundary state $\theta_B = \underline{\theta}$, then $S$ is indifferent between pooling with $[\underline{\theta}, \theta_1]$ and reporting $\overline{\theta}$ for $\underline{\theta}$ versus separating at $\underline{\theta}$. The existence of SLAPH requires (\ref{theorem_connect})-(\ref{sufficient_cond}) to be jointly satisfied. 
The following corollary summarizes the expected gain of $R$ from investigations at SLAPH.
\begin{corollary}\label{corollary_gain}
At SLAPH, the expected total gain $\Delta^{P}$ produced by the evidences $\{e_j\}_{j=0}^{K-1}$ is given by
\begin{equation}
\Delta^P = \sum_{j=0}^{K-1} \Delta_{j,j+1}.
\end{equation}
\end{corollary}
%

\section{Numerical Example}\label{numerical}

 \begin{figure}
 \includegraphics[width=0.5\textwidth]{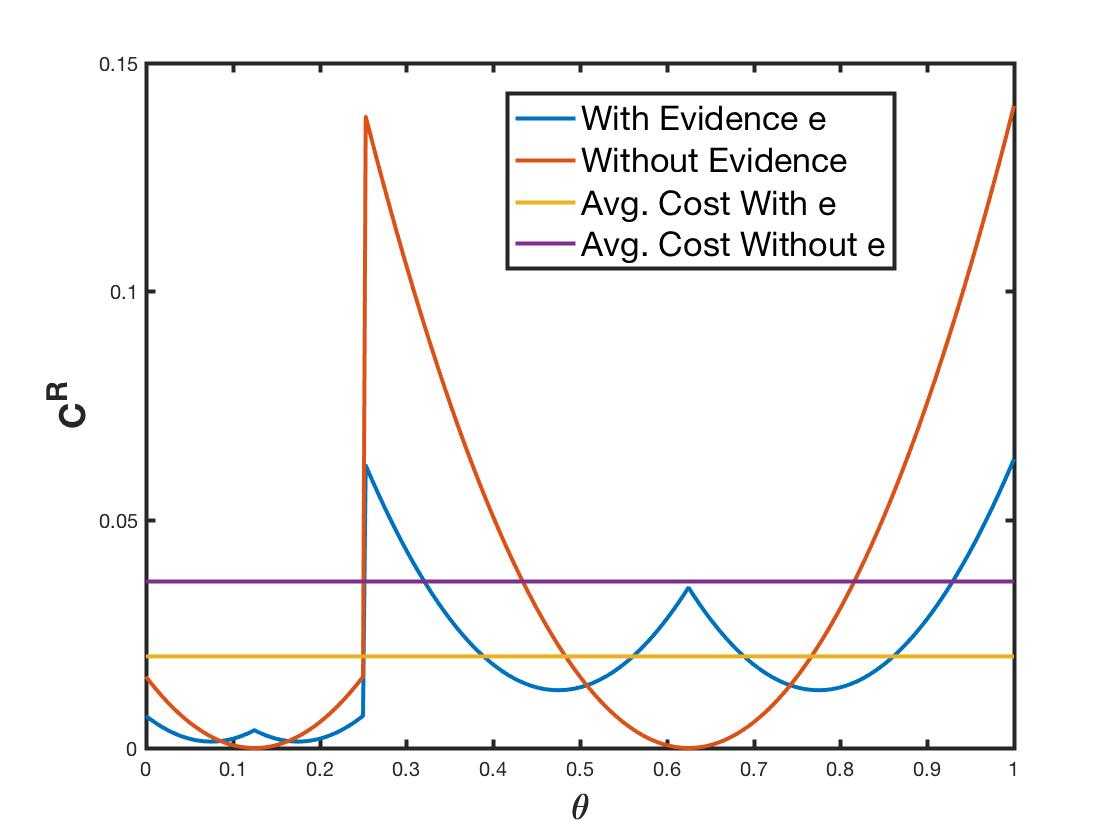}
   \caption{The behavior of costs of $R$ in the example of Fig. \ref{example_multi_pool}, with and without evidence; true positive rates $x=y=0.9$. }\label{cost_R_compare}
  \centering
\end{figure}

In this section, we numerically illustrate the equilibrium behaviors of both $S$ and $R$. In the experiments, we set the minimum value and the maximum value as $\underline{\theta} = 0$ and $\overline{\theta} = 1$, respectively. We consider the case when $\theta$ is uniformly distributed over $[0,1]$, i.e., $F(\theta) = \theta$.
Then, $\hat{a}^R(\theta^{''}, \theta^{'}) = \frac{\theta^{''}+ \theta^{'}  }{2}$.
%
Given the initial condition $\sigma^{S*}(0) = 0$, the differential equation Eq. (\ref{UR_de}) can be solved as 
\begin{equation}\label{solution_SPBNE}
\exp(-\frac{k}{b} \sigma^{S*}(\theta))+ \frac{k}{b} \big( \sigma^{S*}(\theta) - \theta  \big) -1 = 0.
\end{equation}
Since $S$ has a well-defined $\sigma^{S*}(\hat{\theta}) = 1$ for the cut-off state, we have
\begin{equation}\label{theta_cutoff}
\hat{\theta} = 1- \frac{b}{k}(1- \exp(-\frac{k}{b})).
\end{equation}

Fig. \ref{hat_vs_kb} shows the cut-off state $\hat{\theta}$ for different values of $\frac{k}{b}$. As can be seen, $\hat{\theta}$ increases and approaches $1$ as $\frac{k}{b}$ increases. This coincides with the intuition that when the deception cost $k$ is cheap (resp. expensive) relative to the conflict of interest $b$, $S$ prefers the pooling (separating) strategy.

Fig. \ref{SP_PBNE} shows the strategy $\sigma^{S*}$ given $b = \frac{1}{8}$ and $k=0.1$. The cut-off state given by Eq. (\ref{theta_cutoff}) is $\hat{\theta} = 0.3117$. As shown in Fig. \ref{SP_PBNE}, the cost-minimizing strategy of $S$ is $\sigma^{S*}$ for all state $\theta\in[0, 0.3117]$; for all $\theta\in(0.3117, 1]$, there is no solution $\sigma^{S*}$ satisfying Eq. (\ref{solution_SPBNE}), and $S$ prefers to play pooling strategy $\sigma^S(\theta) = 1$.
Fig. \ref{sigma_vs_theta} illustrates the change of $\sigma^{S}$ when $b$ is fixed and $k$ changes.

Fig. \ref{example_multi_pool} shows an example of multiple-pool P-PBNE with $b=\frac{1}{8}$, $k = \frac{1}{64}$. In this example, $\hat{\theta} = 0.06$ and $\theta_B = 0$. Suppose the belief $\mu^R(\theta|m)$ informs $R$ that $\theta\in \Theta_{0,1}$, and $R$ partitions $\Theta_{0,1}$ into two investigation regions, $\Theta^0_{0,1}$ and $\Theta^1_{0,1}$, by setting $\theta^{I}_{0, 1} = \hat{a}(0, 1) = \frac{1}{8}$. 
If $e= 0$, $\sigma^{S*,p} = a^0_{0,1} = \frac{1}{16}$; otherwise, $\sigma^{S*,p} = a^1_{0,1} = \frac{3}{16}$.
Here, $\bar{a}(0,1) = \sum_{i=0}^1P(\Psi^i)a^i_{0,1} = \frac{1}{8}$ and $\bar{a}(1,2) = \sum_{i=0}^1P(\Psi^i)a^i_{1,2} = \frac{5}{8}$. Then, $C^A(\frac{1}{8}, \frac{1}{4}) = C^A(\frac{5}{8}, \frac{1}{4})$ and $C^S(\frac{1}{8}, 0, 1) \geq C^S(0, 0, 0)$. Therefore, a SLAPH exists for this example.
Fig. \ref{cost_R_compare} illustrates the corresponding behavior of $C^R$ when the true positive rates are $x=y=0.9$. The expected cost with evidence (expected $C^R=0.02$) is less than the expected cost without evidence (expected $C^R=0.04$).

%
%
\section{Conclusion}\label{conclusion}

Deception is a technique that can be viewed as an approach for attacks as well as defense. Understanding deception quantitatively is pivotal to provide rigor, predictability, and design principles.
In this work, we have studied deceptions over a continuous one-dimensional information space. We have modeled the strategic interactions between the deceiver and the deceivee by a signal game in which the deceivee uses her prior and posterior beliefs and the evidence acquired through investigations to make her best-effort decision while the deceiver strategically manipulates the information to mislead the deceivee.
We have studied the perfect Bayesian Nash equilibrium (PBNE) as the solution concept to analyze the outcome of the deception game and characterize the deceivability of the game. 
We have shown that the deception game admits a class of PBNE called SLAPH (Separating in Low states And Pooling in High states). 

Our results have shown that there is a unique undeceivable region if the deception cost is sufficiently large compared to the conflict of interest between the deceiver and the deceivee. However, a full undeceivable region does not exist and there exists a deceivable  region. We have also shown that the deceivable region can be partitioned into multple sub-deceivable regions without increasing total costs for the deceiver when the conflict of interest is insignificant.
We have identified the conditions under which the deceivee can obtain more information for deceivable regions by acquiring evidence while inducing disincentive for the deceiver to deviate from the SLAPH.

Future work could focus on the mechanism design for security systems.
Game theory plays an essential role in mechanism design. Our equilibria as the solution concept can provide guidance on security system design for defense mechanism against deception as well as defense mechanism using deception as an approach. For example, from the deceivee's point of view, the identification of undeceivable regions and deceivable regions can provide additional information for optimal resource allocation to maintain the undeceivable regions while reducing the possible damages in the deceivable regions.

\bibliographystyle{IEEEtran}
\bibliography{deception_quant}

\begin{thebibliography}{10}
\providecommand{\url}[1]{#1}
\csname url@samestyle\endcsname
\providecommand{\newblock}{\relax}
\providecommand{\bibinfo}[2]{#2}
\providecommand{\BIBentrySTDinterwordspacing}{\spaceskip=0pt\relax}
\providecommand{\BIBentryALTinterwordstretchfactor}{4}
\providecommand{\BIBentryALTinterwordspacing}{\spaceskip=\fontdimen2\font plus
\BIBentryALTinterwordstretchfactor\fontdimen3\font minus
  \fontdimen4\font\relax}
\providecommand{\BIBforeignlanguage}[2]{{%
\expandafter\ifx\csname l@#1\endcsname\relax
\typeout{** WARNING: IEEEtran.bst: No hyphenation pattern has been}%
\typeout{** loaded for the language `#1'. Using the pattern for}%
\typeout{** the default language instead.}%
\else
\language=\csname l@#1\endcsname
\fi
#2}}
\providecommand{\BIBdecl}{\relax}
\BIBdecl

\bibitem{cott1940adaptive}
H.~B. Cott, \emph{Adaptive coloration in animals}.\hskip 1em plus 0.5em minus
  0.4em\relax Methuen; London, 1940.

\bibitem{vrij2008increasing}
A.~Vrij, S.~A. Mann, R.~P. Fisher, S.~Leal, R.~Milne, and R.~Bull, ``Increasing
  cognitive load to facilitate lie detection: The benefit of recalling an event
  in reverse order,'' \emph{Law and human behavior}, vol.~32, no.~3, pp.
  253--265, 2008.

\bibitem{gneezy2005deception}
U.~Gneezy, ``Deception: The role of consequences,'' \emph{American Economic
  Review}, vol.~95, no.~1, pp. 384--394, 2005.

\bibitem{bodmer2012reverse}
S.~Bodmer, D.~M. Kilger, G.~Carpenter, J.~Jones, and J.~Jones, \emph{Reverse
  deception: organized cyber threat counter-exploitation}.\hskip 1em plus 0.5em
  minus 0.4em\relax McGraw-Hill New York, 2012.

\bibitem{powell2007allocating}
R.~Powell, ``Allocating defensive resources with private information about
  vulnerability,'' \emph{American Political Science Review}, vol. 101, no.~4,
  pp. 799--809, 2007.

\bibitem{brown2005two}
G.~Brown, M.~Carlyle, D.~Diehl, J.~Kline, and K.~Wood, ``A two-sided
  optimization for theater ballistic missile defense,'' \emph{Operations
  research}, vol.~53, no.~5, pp. 745--763, 2005.

\bibitem{pawlickanalysis}
J.~Pawlick, E.~Colbert, and Q.~Zhu, ``Analysis of leaky deception for network
  security using signaling games with evidence.''

\bibitem{crawford1982strategic}
V.~P. Crawford and J.~Sobel, ``Strategic information transmission,''
  \emph{Econometrica: Journal of the Econometric Society}, pp. 1431--1451,
  1982.

\bibitem{zhang2017strategic}
T.~Zhang and Q.~Zhu, ``Strategic defense against deceptive civilian gps
  spoofing of unmanned aerial vehicles,'' in \emph{International Conference on
  Decision and Game Theory for Security}.\hskip 1em plus 0.5em minus
  0.4em\relax Springer, 2017, pp. 213--233.

\bibitem{ettinger2010theory}
D.~Ettinger and P.~Jehiel, ``A theory of deception,'' \emph{American Economic
  Journal: Microeconomics}, vol.~2, no.~1, pp. 1--20, 2010.

\bibitem{osborne1994course}
M.~J. Osborne and A.~Rubinstein, \emph{A course in game theory}.\hskip 1em plus
  0.5em minus 0.4em\relax MIT press, 1994.

\end{thebibliography}

%




\end{document}